\documentstyle[preprint,pra,aps]{revtex}
\begin{document}
\draft

\title{Phase Coexistence of a Stockmayer Fluid in an Applied Field}
\author{Mark J. Stevens and Gary S. Grest}
\address{
Corporate Research Science Laboratories,
Exxon Research and Engineering Company,
Annandale, NJ 08801
}
\date{\today}
\protect\maketitle
\protect\widetext
\protect\begin{abstract}
We examine two aspects of Stockmayer fluids which consists of point
dipoles that additionally interact via an attractive Lennard-Jones potential.
We perform Monte Carlo simulations to examine the effect of an applied field
on the liquid-gas phase coexistence and show that a magnetic fluid phase
does exist in the absence of an applied field.
As part of the search for the magnetic fluid phase, we
perform Gibbs ensemble simulations to determine phase coexistence curves at
large dipole moments, $\mu$.
The critical temperature is found to depend linearly on $\mu^2$ for
intermediate
values of $\mu$ beyond the initial nonlinear behavior near $\mu=0$ and
less than the $\mu$ where no liquid-gas phase coexistence has been found.
For phase coexistence in an applied field,
the critical temperatures as a function of the applied field
for two different $\mu$ are mapped onto a single curve.
The critical densities hardly change as a function of applied field.
We also verify that in an applied field
the liquid droplets within the two phase coexistence region
become elongated in the direction of the field.
\protect\end{abstract}
\protect\pacs{}

\narrowtext

\section{Introduction}

Dipolar fluids have a simple, but anisotropic, long-range pair potential which
presents an interesting new set of problems for statistical physics.
Manifestations of dipolar fluids include ferrofluids (FF), magnetorheological
fluids (MR), electrorheological fluids (ER) and polar fluids
\cite{Rosensweig85,Halsey92,Halsey93a,Leeuwen94b}.
These fluids have great technological promise and in some cases applications,
as they possess the dual properties of being fluid and magnetic
(electric).
In most cases, an applied field is present.
For the MR and ER fluids, an applied field is essential, since only
induced dipoles exists.


Simulations of simple dipolar fluids have discovered several
interesting phenomena.
The fluid can become magnetized in the absence of an applied field at high
densities \cite{Wei92a,Wei92b}.
For the simplest model of hard or soft sphere dipoles in zero field,
no liquid-gas phase coexistence has been found in contrast to predictions of
mean field theory \cite{Caillol93,Leeuwen93b,Stevens94}.
In an applied field phase coexistence does occur for this system,
yet it is not the usual
gas-liquid or even isotropic-magnetic liquid phase coexistence
\cite{Stevens94}.
Coexistence in the absence of an applied field does not occur in part because
the only attractive interaction is the anisotropic dipolar interaction which
tends to produces chains instead of droplets.
Phase coexistence does occur in zero field if a sufficiently strong,
short-range
attractive interaction is added to the dipolar interaction \cite{Leeuwen93b}.

One model that exhibits two phase coexistence in zero external field
is the Stockmayer fluid (SF)
\cite{Shell,Pollock80}.
This system consists of long-range dipoles that have an additional short-range
Lennard-Jones (LJ) interaction.
Since the LJ interaction alone is sufficient to produce a gas-liquid
coexistence
\cite{Hansen90}, it is clear that the Stockmayer fluid must also have a
gas-liquid coexistence at least for small dipole moments.
Recent simulations \cite{Shell} have calculated
the coexistence curves for several nonperturbative values of the dipolar
moment, $\mu$, in zero external magnetic field.
For this reason the Stockmayer fluid is a good system to study the effects of
an applied field on the gas-liquid phase coexistence.

One intriguing aspect of dipolar fluids is the existence of a magnetic
fluid phase in the absence of an applied field.
For soft sphere dipoles, simulations have found a magnetic fluid phase
\cite{Wei92a,Wei92b} at high densities.
Early simulations on the Stockmayer fluid for relatively small $\mu$ found
negative pressures at these high densities \cite{Pollock80}.
Consequently, later studies of the ordering in the fluid
preferred the hard or soft sphere dipolar system \cite{Wei92a,Wei92b}.
We now know that the negative pressures occur because of the two phase
coexistence present in the Stockmayer fluid.
Prior to this work, there have been no simulations to examine whether the
magnetic liquid phase exists for a Stockmayer fluid at densities above the
liquid coexistence curve.
We report here results of simulations for the Stockmayer fluid that
search for the magnetic liquid phase.
The phase coexistence curves were calculated for larger values of
$\mu$ where the magnetic liquid phase is more likely.
Using these calculated coexistence curves we examined the regime above
the liquid coexistence densities and find the magnetic liquid phase.

The Stockmayer fluid provides a simple model system for ferrofluids
\cite{Russier94}.
Recent work \cite{Leeuwen93b,Stevens94,Stevens94b} has shown that
a dipole with a purely repulsive core potential does not model
some aspects of ferrofluids and that the presence of an additional attractive
interaction is essential for phase coexistence.
Experiments on hydrocarbon based ferrofluids have shown that
phase coexistence depends strongly on the solvent \cite{Rosensweig92,Wang94}.
For some solvents phase coexistence in the absence of an applied field does
occur \cite{Wang94}, while for others it does not \cite{Rosensweig92}.
Given that the same behavior occurs in simple dipolar systems as a function
of the strength of the attractive part of the interaction,
an examination of such systems is warranted.

In the next section, we describe the simulation methods employed.
We also show that the variant of the Stockmayer potential used in Ref.
\cite{Leeuwen93b} can be mapped to the standard Stockmayer potential.
In section \ref{sec:coex}, we present calculations of the coexistence
curves for large $\mu$ and verify the mapping determined in section
\ref{sec:method}.
The existence of the magnetic fluid phase is demonstrated in section
\ref{sec:mf}.
In section \ref{sec:coex_h} we examine the effect of an applied field on
phase coexistence.

\section{Simulation Methods}
\label{sec:method}

The dipole-dipole interaction between particles $i$ and $j$ is
\begin{equation}
U_{dd}({\bf r}_{ij}) = \frac{\mu^2}{r^3_{ij}} \left(
	\hat{\mbox{\boldmath $\mu$}}_i \cdot \hat{\mbox{\boldmath $\mu$}}_j -
	3 (\hat{\mbox{\boldmath $\mu$}}_i \cdot \hat {\bf r}_{ij})
	  (\hat{\mbox{\boldmath $\mu$}}_j \cdot \hat {\bf r}_{ij}) \right),
\end{equation}
where ${\mbox{\boldmath $\mu$}}_i$ is the dipole moment of the $i$th
particle, ${\bf r}_{ij}$ is the displacement vector between the two
particles and a hat signifies a unit vector.
There are three relevant parameters for the dipolar systems:
the density, $\rho$,
the dimensionless dipolar coupling strength given by the ratio,
$\lambda = {\mu^2}/{\sigma^3 T}$, where $\sigma$ is the particle diameter,
and the dimensionless applied field, $\eta = {\mu H}/{T}$.
The temperature may also be an independent variable depending on the
nondipolar interactions.
In any case, the dimensionless temperature is $\tau =1/\lambda$.

In addition to the dipole interaction the Stockmayer fluid possesses
a Lennard-Jones pair potential which models the van der Waals interactions
in ferrofluids \cite{Rosensweig85},
\begin{equation}
U_{LJ}(r_{ij}) = 4\varepsilon \left[
			\left( \frac{\sigma}{r_{ij}} \right)^{12}
		       -\left( \frac{\sigma}{r_{ij}} \right)^{6} \right].
\end{equation}
The temperature and $\lambda$ are now independent quantities.
The LJ system with no dipole interaction ($\mu=0$) has a critical point at
$T_c^* = 1.316$ and $\rho_c^*=0.304 $ \cite{Smit90b}, where
variables are given in  reduced form: $T^* = T/\varepsilon$,
$\rho^*=\rho \sigma^3$, $\mu^{*2}= \mu^2/\varepsilon \sigma^3$ and
$H^* =H\sqrt{\sigma^3/\varepsilon}$.
This critical point value is for a LJ cutoff equal to half
the box length which is shifted slightly
from the critical point of the LJ with no cutoff \cite{Nicolas79}.
To maintain consistency we always use a cutoff equal to half the box length
as was done in previous simulations for the Stockmayer fluid in zero field for
several values of $\mu^{*2} \le 6$
\cite{Shell};
at $\mu^*=2$, the critical point is $T_c^*= 2.09$ and
$\rho_c^*=0.289$ \cite{Shell}.

Recent simulations examined ostensibly a different system in which the
strength of the attractive part of the LJ interaction was varied
\cite{Leeuwen93b}.
Specifically, they used the potential
\begin{equation}
U_{LJ6}(r_{ij}) = 4\varepsilon \left[
                \left( \frac{\sigma}{r_{ij}} \right)^{12}
                -\epsilon_6 \left( \frac{\sigma}{r_{ij}} \right)^{6} \right],
\end{equation}
where $\epsilon_6$ is the constant used to vary the strength
of the attractive interaction.
We will refer to the system of dipoles with $U_{LJ6}$ as the SF6 system.
They found phase coexistence for $\epsilon_6 \agt 0.30$.
For smaller values of $\epsilon_6$, two phase coexistence was not detected by
the Gibbs ensemble simulations and near the expected critical points
chain formation was found instead.
Thus, there seems to be a minimum amount of attraction necessary for a
dipolar liquid phase to exist.
This is consistent with the absence of phase coexistence for the soft
sphere dipolar system, $\epsilon_6=0$ \cite{Stevens94,Stevens94b}.

The significant feature of the interaction is not the absolute strength of the
nondipolar attraction, but the relative strength in comparison with dipolar
interaction.
This is because the dipole interaction induces a condensed phase different
from the LJ interaction, and consequently, the phase diagrams for LJ and
strong dipolar (e.g. soft sphere dipoles) systems are different
\cite{Stevens94}.
The dipolar interaction because of its orientation dependence prefers
to aggregate particles in anisotropic chain structures instead of
isotropic droplets like the LJ interaction prefers.
That the relative strengths are the essential quantity is evident from
that fact that $U_{LJ6}$ can be mapped onto $U_{LJ}$ \cite{Stell72}.
Mapping
\begin{equation}
 \sigma' = \sigma/\epsilon_6^{1/6},
\label{eq:sigma}
\end{equation}
and
\begin{equation}
 \varepsilon' = \varepsilon \epsilon_6^2
\label{eq:epsilon}
\end{equation}
converts the LJ6 parameters, $\sigma$ and $\varepsilon$, into LJ parameters
$\sigma'$ and $\varepsilon'$.
The density and temperature are mapped in the following manner
\begin{equation}
 \rho' = \rho/\epsilon_6^{1/2},
\label{eq:rho}
\end{equation}
and
\begin{equation}
 T' = T / \epsilon_6^2.
\label{eq:T}
\end{equation}
We can map the dipolar system, SF6, to SF by
\begin{equation}
 \mu' = \mu / \epsilon_6^{3/4}.
\label{eq:mu}
\end{equation}
Thus, reducing $\epsilon_6$ as was done in Ref. \cite{Leeuwen93b} is equivalent
to increasing the effective dipole moment.
We confirm this mapping in section \ref{sec:coex}.
{}From the results of ref. \cite{Leeuwen93b}, this mapping implies that
for the Stockmayer fluid there is no phase coexistence for
$\mu^* \agt 5$.

Our simulations methods follow that of previous works on dipolar systems
\cite{Wei92a,Wei92b,Shell,Kusalik90,Kusalik94}.
We performed simulations in zero field at $\mu^*=2.5$, 3.0, 3.5 and 4.0.
Simulations as a function of the applied field were performed at
$\mu^* =1$ and 2.5.
The Ewald sum is used to evaluate the dipole interaction in periodic boundary
conditions with the convergence parameter, $\alpha=  5.75$, and reciprocal
vectors are summed to $10\pi/L$ where $L$ is the length of the simulation cell.
The Ewald sum contains a  boundary term for the total potential energy
\begin{equation}
\frac{4\pi}{(2\mu_{BC} +1)L^3} {\bf M^2},
\end{equation}
where ${\bf M} = \sum_i {\bf \mu}_i$ is the magnetization.
We use $\mu_{BC} = \infty$ making the above term zero.
This allows the uniformly magnetized state to occur.
A value of $\mu_{BC}=1$ will lead to  magnetic domains because this
condition prevents the system from having a net magnetization \cite{Wei92b}.
This form of the Ewald sum treats a periodic lattice enclosed in a large
sphere or cube \cite{Deleeuw80}.
These boundary conditions are consistent with
experiments performed between parallel plates.
As mentioned above,
the LJ cutoff was set at half the box length and the usual long range
corrections were employed \cite{Allen87}.

In recent density functional calculations \cite{Groh94}, it was found
that the phase coexistence depends on shape of the system volume.
In these calculations, the system was taken as an ellipsoid of revolution
with axis of revolution $k$ times the other two axes.
The magnetic fluid phase is favored for large $k$.
Simulations with the Ewald sum usually treat a cubic, or equivalently
spherical, system \cite{Deleeuw80}.
The cube has sidelength, $S \gg L$, where $L$ is the simulation box length.
The simulation is generally viewed as modeling a microscopic piece of
the bulk system.
For example, with respect to ferrofluids which are often experimentally
studied confined between parallel plates, the simulation models a cube
within the middle of the system.
Ideally, one wants to choose $\mu_{BC}$ equal to the permeability of
the bulk system, $\mu_{bulk}$,
but that requires {\em a priori} knowledge which one does not have.
Recent simulations have shown that the difference between using
$\mu_{BC}=\infty$ and $\mu_{BC}=\mu_{bulk}$ is small as long as
$\mu_{bulk}$ is large \cite{Wei92b}.

To examine the liquid structure we performed constant volume canonical
ensemble simulations with $N=256$ particles.
These simulations ran for a least $10^5$ MC cycles with
each cycle comprised of an attempt to translate and rotate each particle.
The particle step size was chosen to achieve about 50\% acceptance.
The rotation step was usually set so that the component of the new moment in
the
direction of the old moment was at most 0.2.
This was done for most cases since the acceptance rate could not be
lowered to 50\%.
The rotation step size was lowered below 0.2, when that would yield
a 50\% acceptance.

Gibbs ensemble simulations were performed to determine the
coexistence curves and critical points  \cite{Smit93,Panagiotopoulos92a}.
Below the critical temperature
the Gibbs ensemble directly gives the two coexistence densities at a
given temperature.
For most of the Gibbs simulations, $N=512$.
Some simulations were performed with $N=1024$ to test the $N$ dependence.
The Gibbs simulations required at least $10^4$ cycles.
The larger $\mu^*$ simulations required runs 2-5 times longer.
Long runs were also performed for temperatures near $T_c$.
Each cycle included an attempt to move each particle once,
100 attempts to change the cell volume and
500 attempts to exchange particles between the two cells.
Only a small percentage of exchanges is accepted in these Gibbs simulations.
Thus, an extremely large number of attempts is needed to gather reasonable
statistics.
For our simulations at least 10,000 accepted exchanges occured in
10,000 cycles.
We have verified that both the pressures and the chemical potentials are the
same in the two cells.
The pressure is calculated from the virial expression and the
chemical potential is calculated from the overlap of the particle
insertion and extraction energy distributions \cite{Powles82}.

Our simulations in the Gibbs ensemble were performed with
$H^*$ ranging from 0 to 5.
The critical temperature, $T_c$, and density, $\rho_c$, were determined by
fitting the calculated coexistence curves using the law of rectilinear
diameters and the usual scaling law for the density with exponent
$\beta=0.32$ \cite{Smit90}.

We characterize the system structure through the order parameter, $P_1$,
defined as
\begin{equation}
P_1 = \frac{1}{N} \sum_{i=1}^N \hat{\mbox{\boldmath $\mu$}}_i
	\cdot {\bf \hat{d}}
	= \frac{1}{\mu N} {\bf M} \cdot {\bf \hat{d}},
\end{equation}
where ${\bf \hat{d}}$ is the director and ${\bf M}$ is the total magnetization
of the system.
For a completely magnetized system, $P_1=1$.

\section{Phase coexistence at large $\mu^*$}
\label{sec:coex}

In order to examine the possibility of a magnetic fluid phase,
we performed simulations at larger $\mu^*$ than previous works
\cite{Shell}.
Using Gibbs ensemble simulations we determined the coexistence curves for
$\mu^*=2.5$, 3.0, 3.5 and 4.0.
The critical points are given in Tables \ref{tab:coex}.
We then performed constant volume canonical ensemble simulations at densities
above the liquid coexistence densities looking for a magnetic fluid phase.
Before discussing the magnetic fluid phase, we discuss
the $\mu^*$ dependence of the critical point.

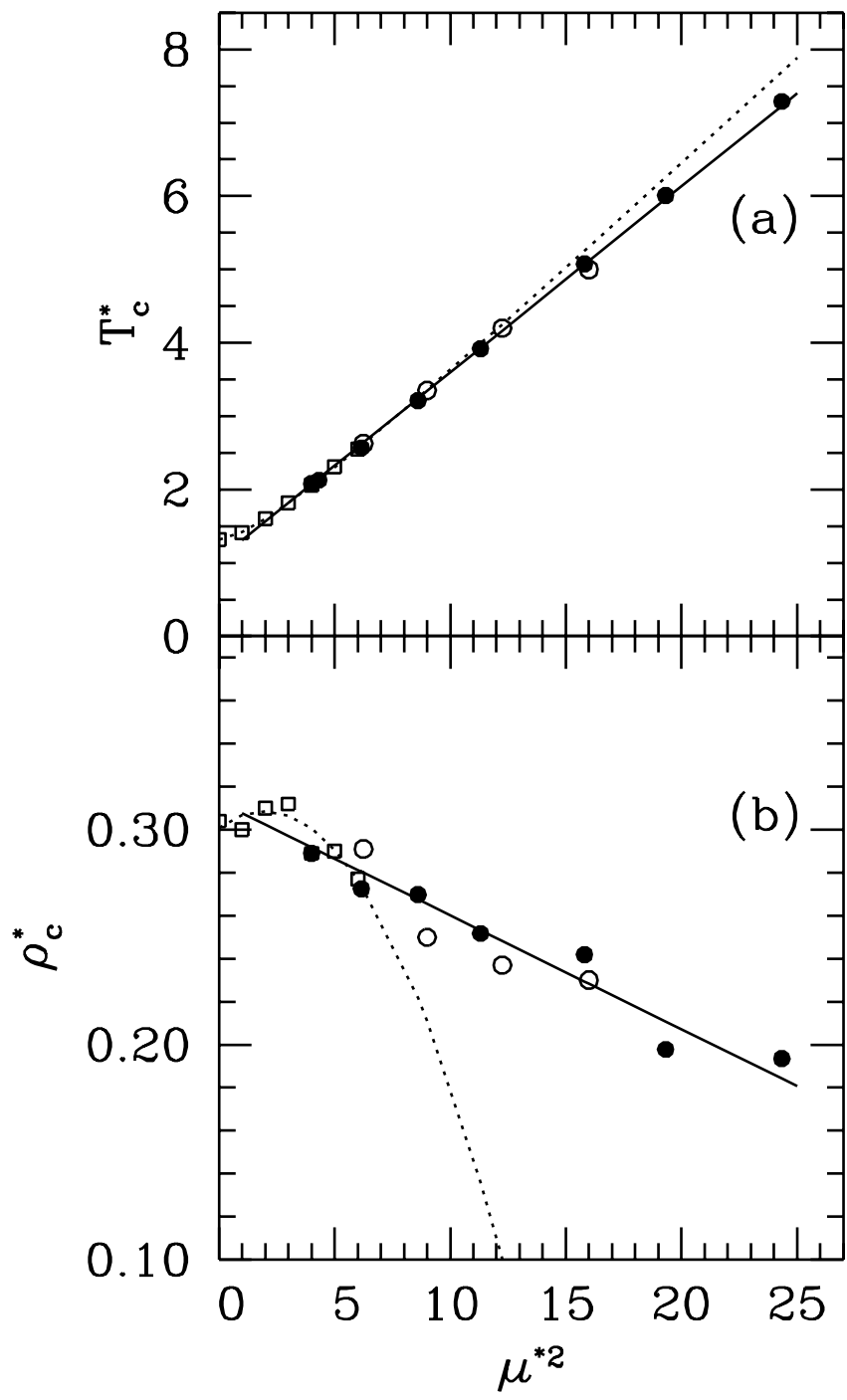
\begin{figure}
\caption{In (a) the critical temperature, $T_c^*$, as a function of
the dipole moment squared, $\mu^{*2}$, and in (b) the critical density
$\rho^*_c$ is plotted.
The open points are from previous works
\protect\cite{Shell}
and the solid points are from this work.
The solid lines are a least squares fit to the nonzero $\mu^*$ data.
The dotted lines are the parameterizations given in ref.
\protect\cite{Shell}(c).
The uncertainty in $T^*_c$ is about $\pm 0.01$ for all data which is smaller
than the points.
}
\label{fig:crit}
\end{figure}

We plot $T_c^*$ as a function of $\mu^{*2}$ in Fig. \ref{fig:crit}(a).
The open squares are from previous simulations
\cite{Shell}, our data are the open circles and
the solid circles are the mapped data of Ref. \cite{Leeuwen93b}.
For $\mu^{*2} \ge 2$ a linear fit to $T_c^*$ with slope 0.254 and intercept
1.06
fits the data well.
This fit must end near $\mu^{*2}\simeq 24$ where no coexistence was found
\cite{Leeuwen93b}.
The parameterization given by van Leeuwen \cite{Shell}(c) fits the $T$ data
except at large $\mu^{*2}$.
Clearly, the data is consistent with the mapping given in
Eqs. \ref{eq:sigma}-\ref{eq:mu}.
For small $\mu^*$, the dipole interaction is effectively a $r^{-6}$ van der
Waals attraction \cite{Stell72}.
This increases the total attraction and consequently, $T_c$ is raised and
$\rho_c$ is lowered as seen in Fig. \ref{fig:crit}.
For large $\mu^*$, this basic trend continues and $T_c$ has a simple
dependence.

Similar results are found for $\rho_c$ in Fig. \ref{fig:crit}(b).
There is more uncertainty in this data, but a least squares fit calculated
as above gives the general decreasing trend with increasing $\mu^*$.
Here, the parameterization of van Leeuwen only works for the original data
fitted.

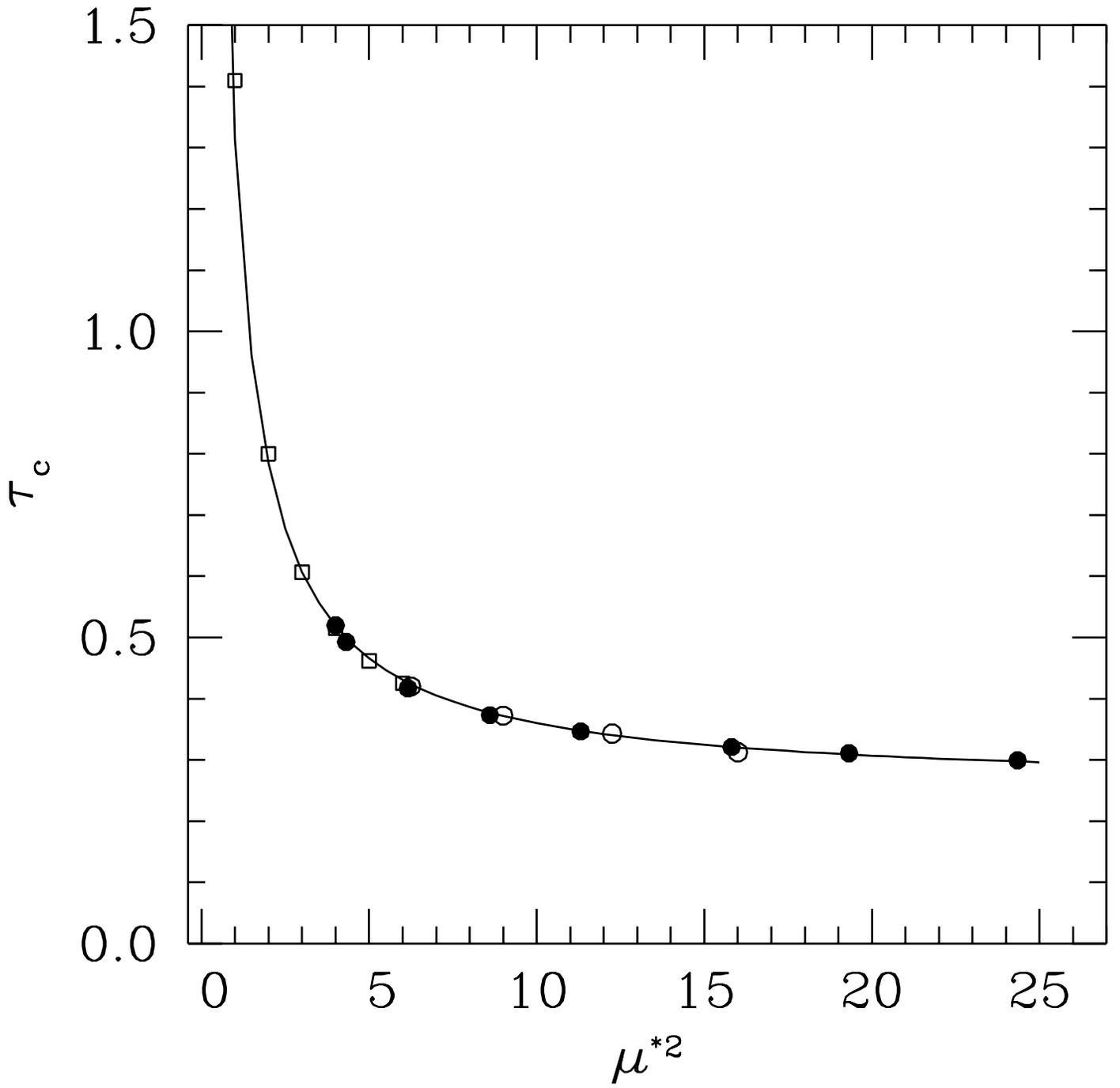
\begin{figure}
\caption{The dimensionless temperature, $\tau_c$, is plotted versus
$\mu^{*2}$ with the same point types as in Fig. \protect\ref{fig:crit}.
The solid line is obtained from the least squares fit in Fig.
\protect\ref{fig:crit}.
}
\label{fig:tau_c}
\end{figure}

As noted earlier, we can also define the dimensionless (dipolar) temperature,
$\tau$.
The critical value, $\tau_c$, is more relevant to determining the dipolar
structure.
In Fig. \ref{fig:tau_c} we plot $\tau_c$ as a function of $\mu^{*2}$.
The solid line is the equation
\begin{equation}
 \tau_c = 0.254 + \frac{1.06}{\mu^{*2}}
\label{eq:taufit}
\end{equation}
derived from the linear fit in Fig. \ref{fig:crit}.
The equation suggests an apparent saturation of $\tau_c$ to about 0.25.
However, the largest value of $\mu^{*2}$ for which phase coexistence has
been found ($\mu^{*2}=24.34$ from $\epsilon_6=0.30$ \cite{Leeuwen93b})
has $\tau_c=0.30$.
Beyond $\mu^{*2} \simeq 24$ Eq. \ref{eq:taufit} does not hold and the
lowest critical dimensionless temperature is $\tau_c=0.30$.
At higher $\mu^*$ the fluid structure in the vicinity of the
$\tau_c$ from Eq. \ref{eq:taufit} exhibits chain formation \cite{Leeuwen93b}.
This change in fluid structure precludes liquid-gas phase coexistence.

\section{Magnetic Fluid Phase}
\label{sec:mf}

We now address the question of the existence of a magnetic fluid phase in
the absence of an applied field in the Stockmayer fluid.
A magnetic fluid phase has already been found in the soft sphere dipolar
system for $\lambda \agt 4$ \cite{Wei92a,Wei92b,Stevens94,Stevens94b}
at high densities.
To determine if the magnetic fluid phase exists for the Stockmayer fluid,
we examine densities larger than the liquid coexistence densities, $\rho_\ell$.
We use the coexistence curves calculated in the last section to determine
the density region of interest.
The coexistence curves for larger $\mu^*$ were calculated in the last section
because a sufficiently large $\lambda \agt 1$ is necessary for existence of the
magnetic fluid phase as found for the soft sphere system.
For $\lambda \alt 1$, thermal interactions dominate and there is no magnetic
fluid phase.

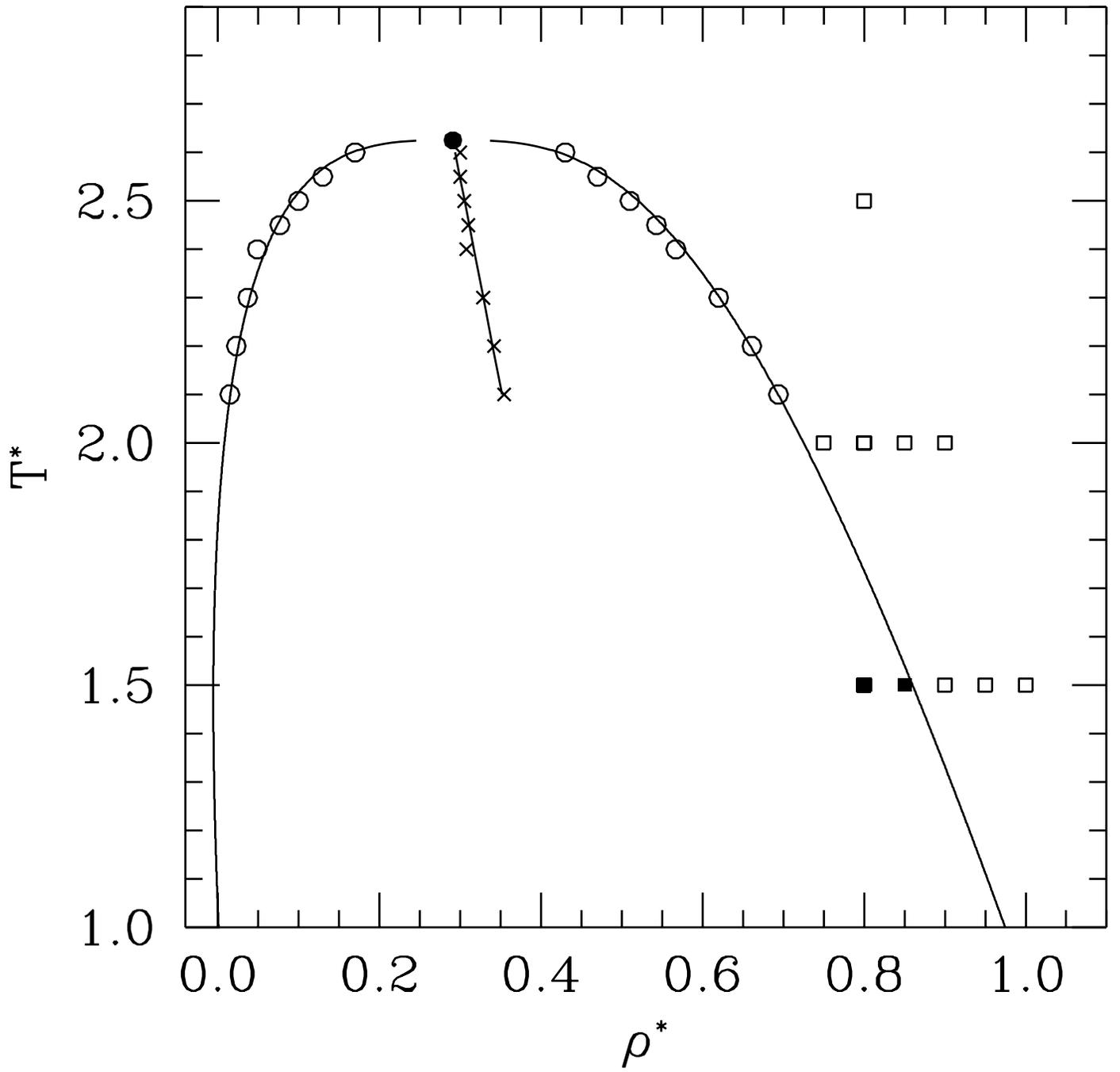
\begin{figure}
\caption{The coexistence curve calculated in the Gibbs ensemble simulation
in the absence of an applied field at $\mu^*=2.5$ is plotted.
The open circles are the coexisting density points found in the simulations.
The solid circle represents the critical point calculated along
with the fitting curve as described in the text.
The square represents points where canonical simulations were performed
to determine the existence of the magnetic fluid phase (see also Table
\protect\ref{tab:liqdat}).
}
\label{fig:coex2.5}
\end{figure}

We find that for the Stockmayer fluid
the structure of liquid coexisting phase in our Gibbs simulations is isotropic.
It is possible for the coexisting liquid phase
to be magnetically ordered \cite{Groh94,Tsebers82,Zhang94,Sano83},
but that is more likely
to occur in the case of a liquid-liquid coexistence, where one liquid phase
is isotropic and the other is magnetic.
In the canonical ensemble simulations at $T << T_c$, we do find some ordering,
but this is most likely a finite size effect.
Figure \ref{fig:coex2.5} shows the Gibbs ensemble data and the fit to the
coexistence curve for $\mu^*=2.5$.
Obviously, far from $T_c^*$, the coexistence curve will probably not follow
the simple fit function used, but we are only
concerned about having a guide to choose where to perform simulations.
The squares show the $\rho^*$ and $T^*$ at which simulations were performed
in this case.
A solid square denotes a negative pressure and is within the coexistence
region as expected.
Similar simulations were performed for $\mu^*=3.0$ and 3.5.
The results are given in Table \ref{tab:liqdat}.

At $T_c$, the values of $\lambda\equiv\lambda_c$ are too low for the magnetic
fluid phase based on our experience with the soft sphere dipolar fluid.
For soft sphere dipoles, we found that for $\lambda=4$, the magnetic fluid
phase occurs close to $\rho^*=1.0$ \cite{Stevens94b}.
For the Stockmayer fluid, we find for $\mu^*=2.5$, $\lambda_c = 2.38$.
Thus, any magnetic fluid behavior, if it occurs at all, will occur at
$T$ much lower than $T_c$.
The situation does not improve much with increasing $\mu^*$ since the value
of $\tau_c = 1/\lambda_c$ (see Fig. \ref{fig:tau_c}) saturates for large
$\mu^*$.
Thus, even for $\mu^*=4.0$, $\lambda_c$ is only 3.17.
To have a chance of finding a magnetic fluid phase, $T$ must be much
lower than $T_c$ and of course, $\rho > \rho_\ell$.
As $T$ decreases, $\rho_\ell$ increases and we will encounter the solid phase
certainly by $\rho^*=\sqrt{2}$, the close packed density.
For hard sphere dipoles, the liquid-solid transition
for $\mu^*=2.5$ occurs at $\rho^*\simeq 1.0$ \cite{Weis93b}.
Thus, the existence of a magnetic fluid phase depends on part on where the
triple point temperature and density are.
If the triple point temperature is sufficiently low, a magnetic fluid phase
may exist in the region above the triple point.

Calculated phase diagrams \cite{Groh94,Zhang94} have shown the critical point
for the isotropic-magnetic coexistence to be at a higher temperature than
the gas-liquid $T_c$ even for $\mu^*=2$.
This is inconsistent with our simulation results, although the density
functional calculations \cite{Groh94} are primarily for ellipsiodal geometries,
not the spherical geometries used in the simulations.
The anisotropy of the ellipsoid promotes the magnetic fluid phase.
For spherical geometry, the results are consistent with our results.
Our simulations also show that the value of $\lambda$ near $T_c$ is too small
for ordering to occur.
Table \ref{tab:liqdat} shows that $P_1$ is small
for $\mu^*=2.5$ at $\rho^*=0.8$ and $T^*=2.5$ which is slightly below $T_c^*$.
Furthermore, as $T^*$ decreases at this density, it is not until
$T^*=1.5$ that $P_1 \ge 0.5$, and at $T^*=2.0$ $P_1 < 1/2$ for $\rho^*<1.0$.
Thus, for the spherical geometry, the simulations show that
the gas-liquid $T_c$ is too high for a magnetic fluid phase to exist above it.

For $\mu^*=2.5$, Table \ref{tab:liqdat} shows that for $T^* \alt 1.5$,
$P_1 >0.5$.
We take the magnetic liquid transition density  to be where  $P_1=0.50$.
There will be of course finite size effects \cite{Binder88} which tend with
increasing $N$ to lower $P_1$ below the transition and increase it above the
transition.
Here, we are mainly interested in the existence of the transition as opposed
to pinning down the transition point which would require simulations with
much larger values of $N$.
For $\lambda=4.19$, we find the transition at $\rho^*\simeq 0.90$.
There is the possibility that the system is a supercooled liquid at this
density.
However, since the transition for
hard sphere dipoles at $\lambda=6.25$  is at $\rho^*\simeq 1.0$,
we are most likely below the liquid-solid transition.
Furthermore, we have found that the fcc crystal phase melts at $\rho^*=1.0$,
although the fcc crystal is most likely not the solid phase ground state
\cite{Weis93b}.
In general, Table \ref{tab:liqdat} shows that for $\lambda \agt 4.0$ and
$\rho^* \agt 0.90$, $P_1 \ge 0.50$ and a magnetic liquid regime exists.

\section{Phase coexistence in an applied field}
\label{sec:coex_h}

Phase coexistence in an applied field was studied at two dipole moments,
$\mu^*=1.0$ and 2.5.
We chose $\mu^*=1$ since at $T_c(H=0)$, $\lambda =0.71$ so that the
thermal and LJ interactions are about of equal strength.
For $\mu^*=2.5$, $\lambda = 2.38$ at $T_c(H=0)$ and the dipole interactions
are significantly larger than the LJ interactions.
This is especially true near $T_c$.
One might expect that the results for the two dipoles strongly differ,
but we find that much of the results can be described in terms of
dimensionless quantities that remove the $\mu^*$ dependence.

For $\mu^*=1$ and large $H^*$, we encountered some difficulties
in obtaining accurate values of the coexisting densities close to the critical
point.
This is a problem that has been observed before \cite{Smit93}.
The free energy surface becomes rather flat and the simulation can become
trapped away from the two minima.
One might expect the problem to be worse for $\mu^*=2.5$, but we found this
not to be the case.

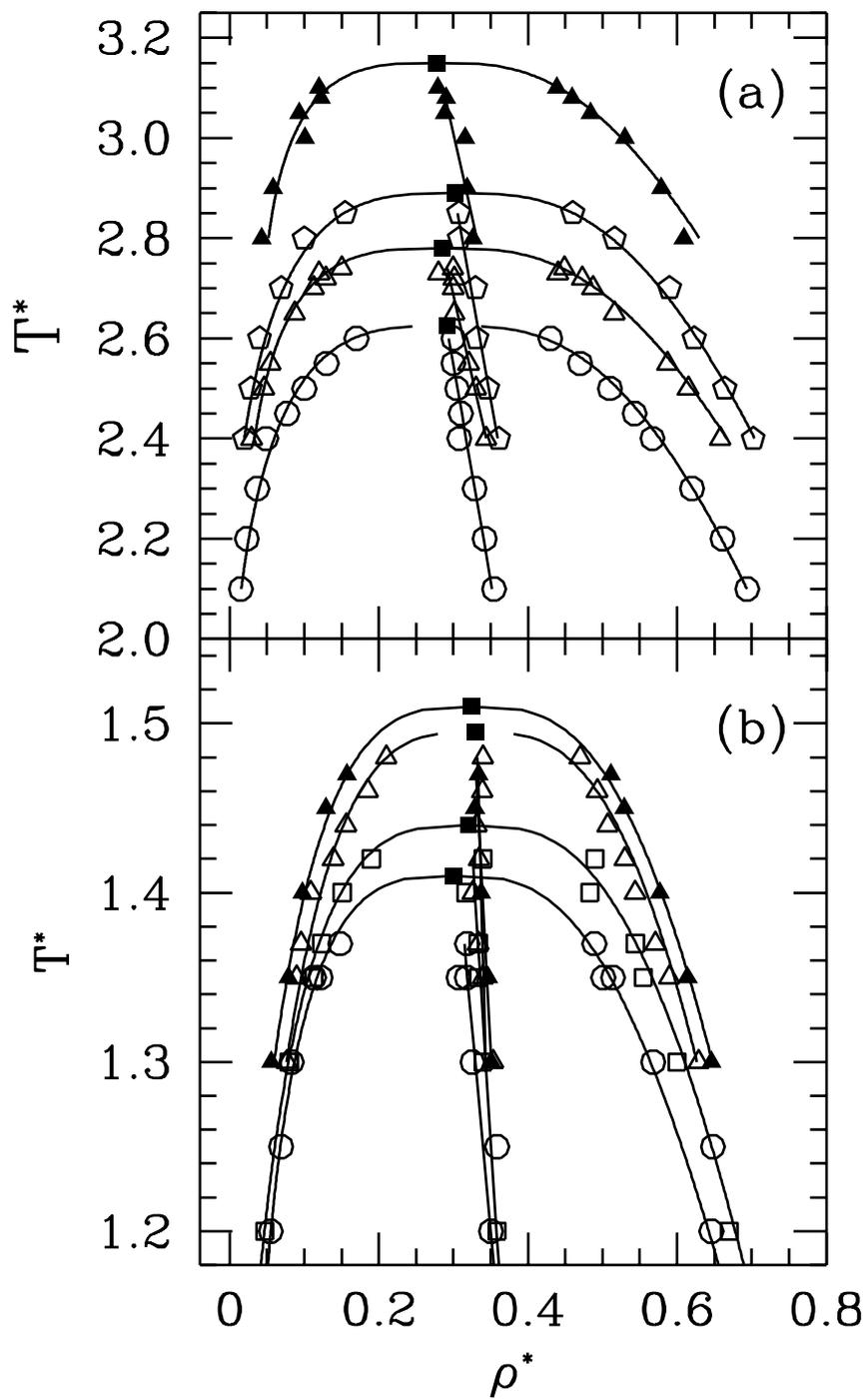
\begin{figure}
\caption{The coexistence curves within an applied field are plotted
(a) for $\mu^*=1$ at $H^*=0$, 1, 2 and 3, and (b) for $\mu^*=2.5$ at $H^*=0$,
0.5, 1.0, 2.0 and 5.0.
}
\label{fig:coex_h}
\end{figure}

Figure \ref{fig:coex_h} shows the phase coexistence curves for $\mu^*=1$ and
$\mu^*=2.5$ at selected fields.
As $H$ increases, the dipoles become progessively aligned with the field
direction.
The dipolar interaction between pairs of particles is then stronger and the
dipole moments are more correlated.
The critical temperature increases with field due to the stronger dipolar
interactions as is found in the absence of an applied field.
In contrast to the varying $T_c$, the critical density changes at most
only slightly.
For $\mu^*=1.0$ the shape of the coexistence curve changes slightly as can
be seen by the fact that the midpoint line becomes almost vertical.
However, we can still fit the curves with $\beta=0.32$ and the $\mu^*=2.5$
midpoint lines maintain a negative slope.
Because the midpoint line was more vertical for $\mu^*=1$ than $\mu^*=2.5$,
more data was needed near $T_c$ which tended to encounter the
convergence problem mentioned above.

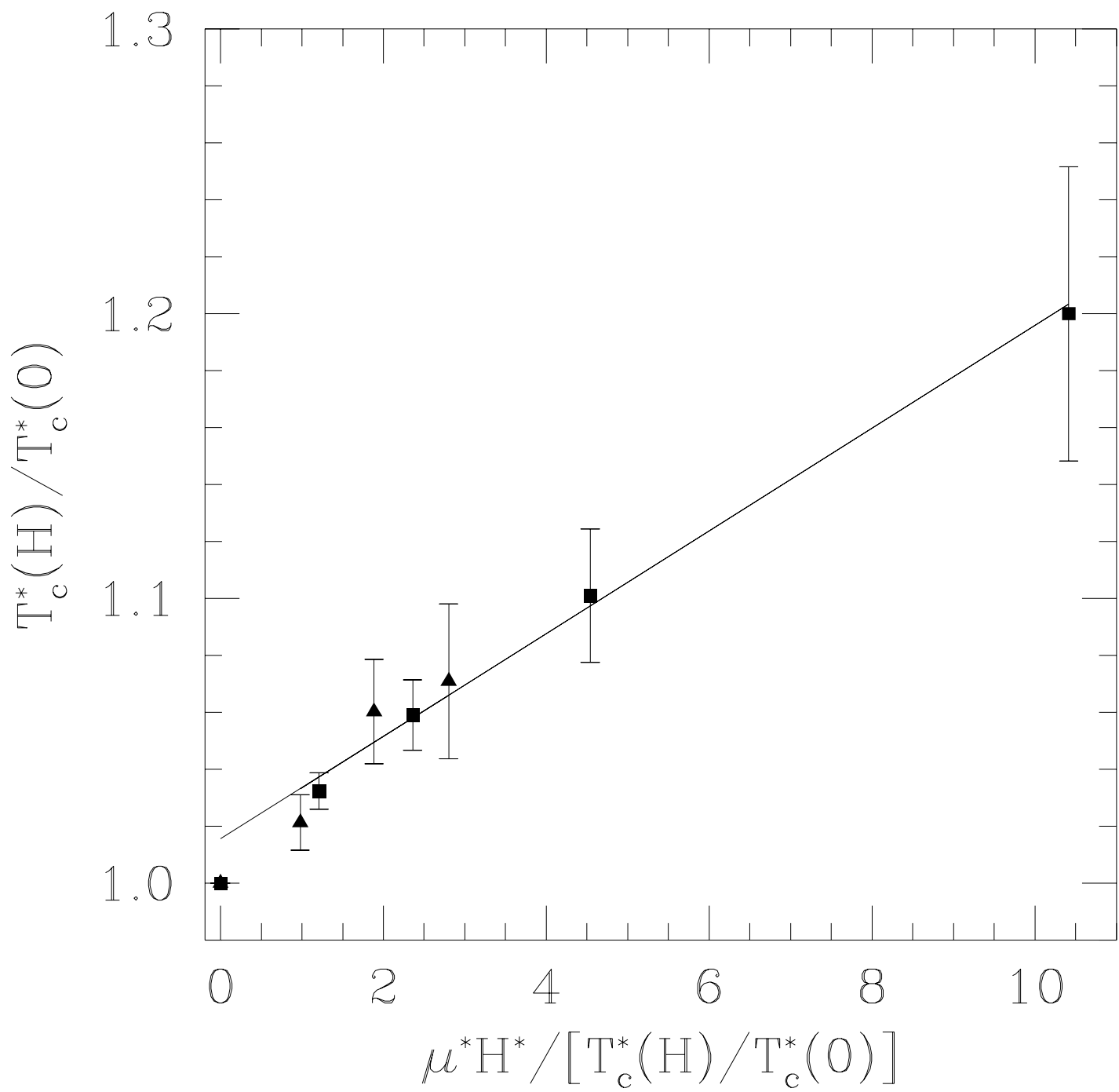
\begin{figure}
\caption{The critical temperature for systems exhibits some scaling when
plotted versus the dimensionless field.
In order for the data of the two $\mu^*$ to be on the same curve, we
scale $T_c^*(H)$ by the zero field $T_c^*$.
This is also done in the dimensionless field.
The solid line is a least squares fit to the nonzero field data.
}
\label{fig:tc_h}
\end{figure}

The field dependence of the critical temperature can be simplified by
examining the critical temperature at $H$, $T_c(H)$, relative to the zero
field critical temperature, $T_c(0)$ in terms of the ratio
\begin{equation}
 T_c^H = \frac{T_c(H)}{T_c(0)}.
\label{eq:t_c_h}
\end{equation}
A collapse of the data (Fig. \ref{fig:tc_h}) for our two dipole moments is
obtained when we plot $T_c^H$ versus the dimensionless field using $T_c^H$ as
the temperature,
\begin{equation}
 \eta_H \equiv \frac{\mu H }{k_BT_c^H}.
\label{eq:eta_H}
\end{equation}
We have drawn a least squares fit to the data excluding the $H=0$ point
which fits the data within the uncertainty.
Near $H=0$ there must be some nonlinear behavior.
At large $H$, i.e. beyond the saturation field, a $\mu$ dependence is expected
because saturation is $\mu$ dependent.
We have not reached the saturation $H$ in our simulations.
In order to determine $T_c$ for the infinite field, we performed
Gibbs ensemble simulations with the dipole moments fixed in the $z$-direction.
For this case, we find $T_c^* = 3.64$ for $\mu^*=2.5$.
This yield $T_c^H=1.38$ which from the fit occurs for $\eta_H=20.5$
which is beyond the range we have studied.
Thus, for the range of $H$ we studied, we see no effects of field
saturation.

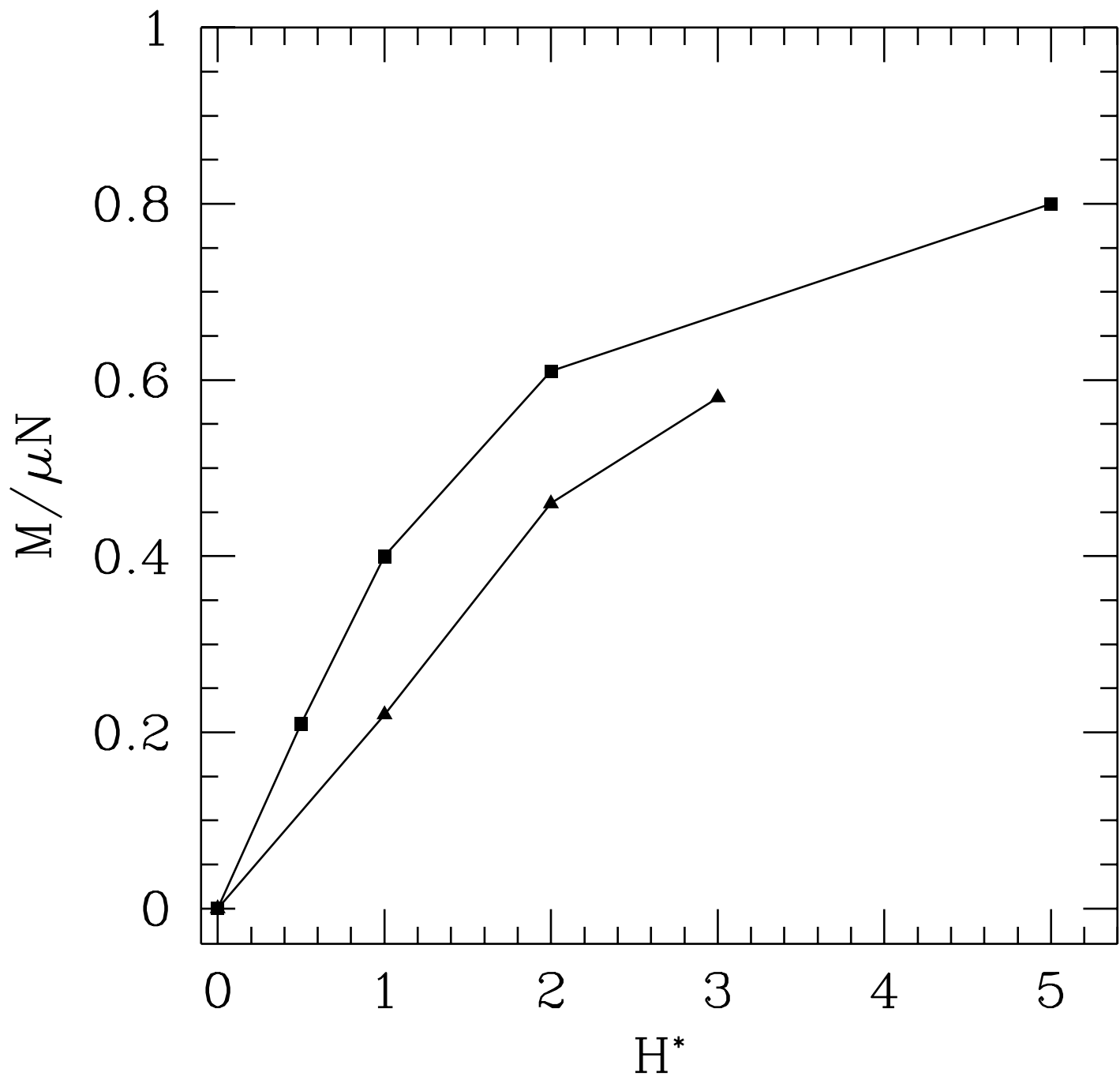
\begin{figure}
\caption{The magnetization per particle versus the applied field
for $\mu^*=1$ (triangles) and $\mu^*=2.5$ (squares).
}
\label{fig:M}
\end{figure}

We can also get an idea of how near the systems are to saturation by
calculating the magnetization, $M$, or equivalently, $P_1$.
In Tables \ref{tab:coex1} and \ref{tab:coex2.5_h} we list the values of the
order parameter, $P_1$.
We want to calculate $M$ at the critical point.
The magnetization in the gas phase is basically constant as the dipoles
only interact weakly.
The interaction energy between a dipole pair will be $k_BT$ at a separation of
$r=\lambda^{1/3}\sigma$.
For $\rho^* \simeq 0.30$, the average separation is
$a=\rho^{-1/3} \simeq 1.5\sigma$.
We are working in the range of $\lambda <4$, which gives $r=1.6\sigma$.
Thus, the dipolar interaction energies are at most equal to $k_BT$.
We take $M_c \equiv M(T_c)$ to be the value of $M$ for the gas phase.
In Fig. \ref{fig:M} we plot $M/\mu N$ for the two cases.
Neither case has reached saturation, as we expected since there is no $\mu$
dependence in our $T_c^H$.

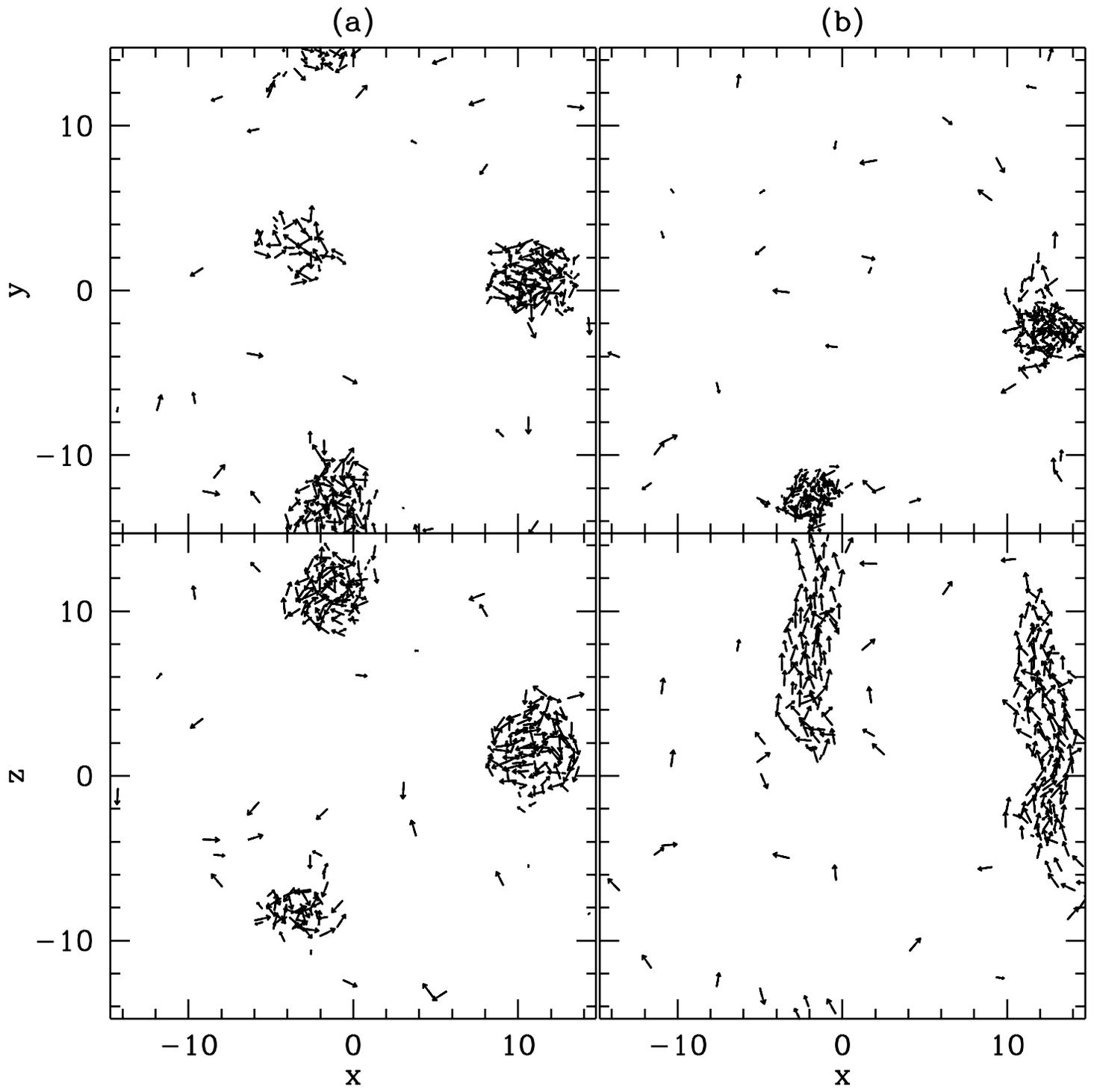
\begin{figure}
\caption{Projection plots for (a) $H^*=0$ and (b) $H^*=1.0$ for
$\mu^*=2$ at $\rho^*=0.1$ and $T^*=1$ show the effect of an applied field
on the liquid droplet shape in the coexistence region.
The field is parallel to the $z$-direction.
}
\label{fig:drops}
\end{figure}

One of the important question concerning dipolar fluids (FF, MR, and ER)
presently under consideration
\cite{Halsey92,Halsey93a,Wang94,Grasselli94,Fermigier92,Bacri82}
is the shape of the liquid coexisting phase.
We can only examine the structure at the particle level and are mainly
concerned with confirming the basic effect of elongation of a
liquid droplet in the applied field.
We have performed a constant volume canonical ensemble simulation for
$\mu^*=2$ at $\rho^*=0.01$, $T^*=1.0$ and $H^*=0$ and 1.0.
In Fig. \ref{fig:drops} we show projection plots for the two different $H^*$.
In the absence of a field, the droplets are spherical on average as expected,
and in the presence of a field the droplets become extended along the
field direction ($z$ in Figure) and in this case two of the droplets at
$H^*=0$ coalesced into one.
Thus, we find the two main effects of an applied field on liquid droplets:
elongation in the field direction and coalescence \cite{Liu93a,Liu93b}.

\section{Conclusion}
\label{sec:concl}

The results of the the present simulations give a better understanding
of the phase diagram of the Stockmayer fluid in both zero or nonzero fields.
The mapping of the SF6 system onto the SF system shows that dipole interaction
strength in comparison with the LJ interaction strength
is the determining quantity for the occurance of phase coexistence.
The mapping is particularly useful as discussed below when comparing with
experimental systems as discussed below.
The $\mu$ dependence of the critical point is rather simple for most
of the range over which SF phase coexistence occurs.
In this range, the temperature, $T_c^*$ depends linearly on $\mu^2$.
Linear behavior is also true for $\rho_c$, although there is more uncertainty
here.
The magnetic fluid phase does appear at high densities for sufficently
large $\mu$.
At least for the spherical geometry, the tricritical temperature appears to be
below $T_c$.
In an applied fields, a $\mu$-independent scaling can be obtained at least
for a broad range of fields for
the critical temperature as a function of applied field when the temperature
is scaled by the zero field $T_c$.

One of the important conclusions of previous works \cite{Stevens94,Stevens94b}
is that the hard or soft sphere dipolar systems were insufficient
as models for ferrofluids (and most likely MR and ER fluids).
Some added central force attraction is required such as found in the SF or SF6
potentials.
Dipole moments in terms of $\lambda$ are about 1, although ferrofluids
possess a large polydispersity \cite{Rosensweig85}.
However, the residual strength of an attractive interaction such as van
der Waals is unknown.
Thus, the value of $\epsilon_6$ must be determined from some experimental data.
If phase coexistence occurs in zero field,
one way of determing $\epsilon_6$ is to map the value of $T_c$ for $H=0$,
and the given value of $\mu$ onto the plot of $T_c$ versus $\epsilon_6$.
In the cases where no phase coexistence occurs at $H=0$, one must
determine $\epsilon_6$ from the value of $T_c$ at $H>0$.
The linear relationships between $T_c$ or $\rho_c$ and $\mu^2$ simplify
this procedure somewhat.
In some of the experiments \cite{Rosensweig92} phase coexistence curves
are not fully measured, but slices in the $H$-$\rho$ plane at fixed $T$
are measured.
To compare with these experiments requires calculation of coexistence
curves not only at several $H$, but also several $T$ and $\epsilon_6$
in order to determine the correct $\epsilon_6$ and the make the correct slices.

The structure at the particle level in ferrofluids has not been
resolved experimentally.
Simulations naturally offer a means to examine this structure.
The structure of the columnar objects \cite{Rosensweig85,Krueger80}
that form in ferrofluids is an open question.
In the simulations of soft sphere dipoles, chains form consisting
of connected single particles as in a polymer \cite{Stevens94,Stevens94b}.
These chains do not aggregate to form the columnar structures as has been
observed experimentally.
In contrast, we find in the simulations of the Stockmayer fluid
that the polymeric chains do not form.
In an applied field, the zero field droplets distort becoming elongated
in the field direction.
Droplets are also seen to coalesce.
This suggests that in ferrofluids that the columns are formed by
initial formation of liquid droplets that are distorted into an elliptical
shape and coalesce in an applied field to form columns spanning the
experimental cell.

\begin{table}
\caption{Critical points.}
\begin{tabular}{d d d d }
$\mu^*$  & $H^*$ & $T_c^*$ & $\rho_c^*$ \\ \hline
2.5& 0.0&  2.63(1)&  0.29(1)  \\
3.0& 0.0&  3.35(1)&  0.25(1)  \\
3.5& 0.0&  4.20(1)&  0.24(1)  \\
4.0& 0.0&  5.07(5)&  0.24(1)  \\ \hline
1.0\tablenotemark[1] & 0.0&  1.41(1)&  0.30(1) \\
1.0& 1.0& 1.44(1)& 0.32(1) \\
1.0& 2.0& 1.49(1)& 0.33(1) \\
1.0& 3.0& 1.51(1)& 0.32(1) \\ \hline
2.5& 0.5& 2.71(1)& 0.285(1) \\
2.5& 1.0& 2.78(1)& 0.285(1) \\
2.5& 2.0& 2.89(1)& 0.302(1) \\
2.5& 5.0& 3.15(1)& 0.278(1) \\
2.5& $\infty$& 3.64(1)& 0.303(1) \\
\end{tabular}
\tablenotetext[1]{Data from Ref. \cite{Leeuwen93b}.}
\label{tab:crit}
\end{table}

\begin{table}
\caption{Phase coexistence data for $\mu^*=3.0$, 3.5 and 4.0}
\begin{tabular}{d d d d d d }
$T^*$  & $\mu^*$ & $\rho_g^*$ & $P_g^*$ & $\rho_\ell^*$ & $P_\ell^*$ \\ \hline
2.70&   3.0     & 0.011(1) & 0.021(1)& 0.68(1)&  0.01(5)  \\
2.85&   3.0     & 0.020(2) & 0.036(3)& 0.63(2)&  0.02(5)  \\
2.95&   3.0     & 0.033(5) & 0.053(3)& 0.59(2)&  0.05(2) \\
3.00&   3.0     & 0.060(9) & 0.057(5)& 0.57(2)&  0.05(6) \\
3.10&   3.0     & 0.060(8) & 0.077(1)& 0.57(2)&  0.09(6) \\
3.20&   3.0     & 0.072(9) & 0.099(5)& 0.46(3)&  0.09(1)  \\
3.25&   3.0     & 0.086(14)& 0.099(2)& 0.43(4)&  0.11(1)  \\
3.30&   3.0     & 0.13(1)  & 0.124(4)& 0.39(2)&  0.12(1)  \\
3.35&   3.0     & 0.21(2)  & 0.148(5)& 0.30(2)&  0.14(2) \\ \hline
3.50&   3.5     & 0.011(1)& 0.025(2)& 0.64(2)& 0.03(9) \\
3.70&   3.5     & 0.017(1)& 0.043(2)& 0.58(2)& 0.03(3) \\
3.80&   3.5     & 0.031(8)& 0.046(6)& 0.54(2)& 0.03(2) \\
3.90&   3.5     & 0.053(6)& 0.070(6)& 0.50(4)& 0.06(3) \\
4.00&   3.5     & 0.06(1)&  0.084(4)& 0.46(2)& 0.09(3) \\
4.05&   3.5     & 0.065(5)& 0.091(9)& 0.45(2)& 0.10(2) \\
4.10&   3.5     & 0.08(2)&  0.101(4)& 0.41(4)& 0.12(2) \\
4.15&   3.5     & 0.10(2)&  0.111(4)& 0.39(1)& 0.11(3) \\ \hline
4.70&   4.0     & 0.03(6)& 0.04(5) & 0.48(8)&  0.01(4) \\
4.80&   4.0     & 0.05(2)& 0.07(2) & 0.53(2)&  0.10(3) \\
4.90&   4.0     & 0.05(1)& 0.072(4)& 0.50(2)&  0.07(4) \\
4.95&   4.0     & 0.07(1)& 0.09(1) & 0.44(2)&  0.08(3) \\
5.00&   4.0     & 0.06(1)& 0.08(1) & 0.41(2)&  0.07(3) \\
5.02&   4.0     & 0.14(2)& 0.08(5) & 0.41(4)&  0.09(2) \\
\end{tabular}
\label{tab:coex}
\end{table}

\begin{table}
\caption{Data for dense liquid phase for $N=256$.}
\begin{tabular}{d d d d d d }
$\mu^*$&$\lambda_B$&$T^*$ &$\rho^*$&$P^*$ &$P_1$  \\ \hline
2.5	& 2.50	&2.5	&0.80	&2.83(3)  &0.08(1)	\\
2.5	& 3.12	&2.0	&0.75	&0.33(1)  &0.09(1)	\\
2.5	& 3.12	&2.0	&0.80	&0.98(4)  &0.11(4)	\\
2.5	& 3.12	&2.0	&0.85	&1.97(3)  &0.13(3)	\\
2.5	& 3.12	&2.0	&0.95	&5.21(2)  &0.21(3)	\\
2.5	& 4.16	&1.5	&0.80	&-0.93(5)  &0.24(2)	\\
2.5	& 4.16	&1.5	&0.85	&-0.41(4)  &0.36(9)	\\
2.5	& 4.16	&1.5	&0.90	&0.36(4)  &0.51(2)	\\
2.5	& 4.16	&1.5	&0.95	&1.57(7)  &0.62(1)	\\
2.5	& 4.16	&1.5	&1.00	&3.26(5)  &0.69(2)	\\ \hline
3.0	& 4.09	&2.2	&0.80	&-0.35(4)  &0.17(10)	\\
3.0	& 4.09	&2.2	&0.90	&1.74(7)  &0.33(5)	\\
3.0	& 4.50	&2.0	&0.90	&0.60(15) &0.55(6)	\\
3.0	& 4.50	&2.00	&1.00	&3.42(9)  &0.71(1)	\\ \hline
3.5	& 4.08	& 3.0	& 0.80	&0.58(5)  &0.20(4)	\\
3.5	& 4.08  & 3.00  & 0.90	&2.70(7)  &0.29(5)   \\
3.5	& 4.45  & 2.75  & 0.80	&0.00(14) &0.22(8)   \\
3.5	& 4.45  & 2.75  & 0.90	&1.62(12) &0.41(10)   \\
3.5	& 4.90  & 2.50  & 0.90	&0.37(17) &0.64(2)   \\
3.5	& 4.90  & 2.50  & 1.00	&2.98(7)  &0.76(2)   \\
\end{tabular}
\label{tab:liqdat}
\end{table}

\begin{table}
\squeezetable
\caption{Phase coexistence data in fields for $\mu^*=1.0$}
\begin{tabular}{d d d d d d d d}
$T^*$ &$H^*$ &$\rho_g^*$ &$P_g^*$ &$P_1^{(g)}$ &$\rho_\ell^*$ &$P_\ell^*$
&$P_1^{(\ell)}$ \\ \hline
1.10& 1.0& 0.025(2)& 0.023(2)& 0.24(4)& 0.724(5)& 0.024(7)& 0.477(3) \\
1.20& 1.0& 0.047(3)& 0.043(3)& 0.21(2)& 0.669(5)& 0.046(7)& 0.43(1) \\
1.30& 1.0& 0.080(5)& 0.071(3)& 0.22(5)& 0.602(5)& 0.072(6)& 0.366(7) \\
1.35& 1.0& 0.113(7)& 0.092(3)& 0.21(7)& 0.554(7)& 0.083(8)& 0.341(7) \\
1.37& 1.0& 0.122(9)& 0.098(4)& 0.22(2)& 0.54(1) & 0.09(2) & 0.31(1) \\
1.40& 1.0& 0.15(1) & 0.11(3) & 0.22(2)& 0.48(2) & 0.100(7)& 0.30(3) \\   \hline
1.30& 2.0& 0.077(6)&  0.069(2)& 0.46(1)& 0.629(6)&  0.067(7) & 0.60(1) \\
1.35& 2.0& 0.090(5)&  0.079(4)& 0.45(2)& 0.590(4)&   0.08(2)  & 0.57(1) \\
1.40& 2.0& 0.109(5)&  0.096(2)& 0.44(1)& 0.544(8)&   0.094(9) & 0.55(1) \\
1.44& 2.0& 0.157(9)&  0.117(2)& 0.45(2)& 0.50(1) &   0.118(6) & 0.53(2) \\
1.46& 2.0& 0.186(8)&  0.129(3)& 0.45(1)& 0.49(2) &   0.132(7) & 0.52(1) \\
1.48& 2.0& 0.21(3) &  0.138(5)& 0.46(1)& 0.45(2) &   0.141(8) & 0.51(1) \\
\hline
1.30& 3.0& 0.056(8) &  0.055(5)& 0.60(1)&  0.646(3) & 0.05(1) & 0.70(1) \\
1.35& 3.0& 0.078(4) &  0.072(2)& 0.60(1)&  0.614(8) & 0.06(1) & 0.69(1) \\
1.40& 3.0& 0.099(5) &  0.088(2)& 0.58(1)&  0.575(17)& 0.08(2) & 0.67(1) \\
1.45& 3.0& 0.132(11)&  0.108(6)& 0.60(1)&  0.529(6) & 0.104(6)& 0.65(1) \\
1.45& 3.0& 0.129(11)&  0.108(6)& 0.58(1)&  0.518(16)& 0.103(9)& 0.65(1) \\
1.47& 3.0& 0.157(6) &  0.120(1)& 0.58(1)&  0.51(2)  & 0.12(2) & 0.65(1) \\
\end{tabular}
\label{tab:coex1}
\end{table}

\begin{table}
\squeezetable
\caption{Phase coexistence data in fields for $\mu^*=2.5$}
\begin{tabular}{d d d d d d d d}
$T^*$ &$H^*$ &$\rho_g^*$ &$P_g^*$ &$P_1^{(g)}$ &$\rho_\ell^*$ &$P_\ell^*$
&$P_1^{(\ell)}$ \\ \hline
2.30& 0.5& 0.03(1)& 0.045(4)& 0.20(1)&   0.66(1)&   0.05(1)& 0.53(2) \\
2.40& 0.5& 0.04(1)& 0.06(1) & 0.16(1)&   0.62(1)&   0.07(2)& 0.50(2) \\
2.50& 0.5& 0.09(1)& 0.092(3)& 0.21(2)&   0.56(1)&   0.09(3)& 0.44(3) \\
2.60& 0.5& 0.16(3)& 0.118(6)& 0.23(2)&   0.49(2)&   0.12(1)& 0.40(1) \\
2.65& 0.5& 0.14(1)& 0.14(1) & 0.20(5)&   0.44(4)&   0.14(2)& 0.36(1) \\
2.70& 0.5& 0.21(5)& 0.16(1) & 0.22(6)&   0.38(5)&   0.16(3)& 0.33(1) \\ \hline
2.40& 1.0& 0.030(5)& 0.048(6) & 0.34(2)& 0.66(1)&  0.05(3)& 0.68(1) \\
2.50& 1.0& 0.046(5)& 0.069(4) & 0.33(4)& 0.61(1)& 0.07(2) & 0.64(1) \\
2.65& 1.0& 0.09(2) & 0.109(4) & 0.38(4)& 0.51(2)& 0.10(4) & 0.59(1) \\
2.70& 1.0& 0.11(2) & 0.127(8) & 0.39(1)& 0.49(3)& 0.136(3)& 0.57(1) \\
2.72& 1.0& 0.13(2) & 0.133(7) & 0.40(2)& 0.47(1)& 0.13(1) & 0.56(2) \\
2.73& 1.0& 0.12(2) & 0.133(10)& 0.39(4)& 0.44(3)& 0.13(1) & 0.54(2) \\
2.74& 1.0& 0.15(2) & 0.137(7) & 0.43(4)& 0.45(1)& 0.14(1) & 0.55(1) \\ \hline
2.40& 2.0& 0.019(1) & 0.034(4) & 0.61(4)&  0.702(3) & 0.03(5)& 0.80(1) \\
2.50& 2.0& 0.028(3) & 0.047(3) & 0.59(3)&  0.664(5) & 0.06(3)& 0.78(1) \\
2.60& 2.0& 0.040(6) & 0.066(6) & 0.59(3)&  0.623(6) & 0.07(6)& 0.77(1) \\
2.70& 2.0& 0.069(11)& 0.093(6) & 0.60(2)&  0.585(13)& 0.10(1)& 0.75(1) \\
2.80& 2.0& 0.10(2)  & 0.118(10)& 0.60(2)&  0.516(6) & 0.13(1)& 0.72(1) \\
2.85& 2.0& 0.13(2)  & 0.136(3) & 0.61(2)&  0.50(2)  & 0.14(1)& 0.72(1) \\
\hline
2.80& 5.0&  0.043(8)& 0.070(5)& 0.82(1)& 0.609(6)& 0.063(5)& 0.86(1) \\
2.90& 5.0&  0.058(6)& 0.091(6)& 0.80(1)& 0.579(8)& 0.09(2) & 0.86(1) \\
3.00& 5.0&  0.10(2) & 0.118(7)& 0.80(1)& 0.54(2) & 0.12(2) & 0.85(1) \\
3.05& 5.0&  0.09(1) & 0.13(1) & 0.80(1)& 0.48(1) & 0.122(8)& 0.86(1) \\
3.08& 5.0&  0.12(2) & 0.14(1) & 0.80(1)& 0.46(3) & 0.14(3) & 0.83(1) \\
3.10& 5.0&  0.12(2) & 0.14(2) & 0.80(1)& 0.44(2) & 0.14(1) & 0.83(1) \\
\end{tabular}
\label{tab:coex2.5_h}
\end{table}

\narrowtext


\begin{thebibliography}{10}

\bibitem{Rosensweig85}
R. Rosensweig, {\em Ferrohydrodynamics} (Cambridge University Press, Cambridge,
  1985).

\bibitem{Halsey92}
T. Halsey, Science {\bf 258},  761  (1992).

\bibitem{Halsey93a}
T. Halsey and J. Martin, Sci. Am. {\bf 269},  58  (1993).

\bibitem{Leeuwen94b}
M. van Leeuwen, Fluid Phase Equilibria {\bf 99},  1  (1994).

\bibitem{Wei92a}
D. Wei and G. Patey, Phys. Rev. Lett. {\bf 68},  2043  (1992).

\bibitem{Wei92b}
D. Wei and G. Patey, Phys. Rev. {\bf A46},  7783  (1992).

\bibitem{Caillol93}
J.-M. Caillol, J. Chem. Phys. {\bf 98},  9835  (1993).

\bibitem{Leeuwen93b}
M. van Leeuwen and B. Smit, Phys. Rev. Lett. {\bf 71},  3991  (1993).

\bibitem{Stevens94}
M.~J. Stevens and G.~S. Grest, Phys. Rev. Lett. {\bf 72},  3686  (1994).

\bibitem{Shell}
B. Smit, C. Williams, and E. Hendriks, Mol. Phys. {\bf 68},  765  (1989);
M. van Leeuwen, B. Smit, and E. Hendriks, Mol. Phys. {\bf 78},  271  (1993);
M. van Leeuwen, Mol. Phys. {\bf 82},  383  (1994).

\bibitem{Pollock80}
E.~L. Pollock and B.~J. Alder, Physica {\bf 102A},  1  (1980).

\bibitem{Hansen90}
J.-P. Hansen and I. McDonald, {\em Theory of Simple Liquids} (Academic Press,
  London, 1990).

\bibitem{Russier94}
V. Russier and M. Douzi, J. Coll. Inter. Sci. {\bf 162},  356  (1994).

\bibitem{Stevens94b}
M.~J. Stevens and G.~S. Grest (unpublished).

\bibitem{Rosensweig92}
R. Rosensweig and J. Popplewell,  in {\em Electromagnetic Forces and
  Applications} (Elsevier Science Publishers, New York, 1992), p.\ 83.

\bibitem{Wang94}
H. Wang {\it et~al.}, Phys. Rev. Lett. {\bf 72},  1929  (1994).

\bibitem{Smit90b}
B. Smit, Ph.D. thesis, Rijksuniversiteit Utrecht, The Netherlands, 1990.

\bibitem{Nicolas79}
J. Nicolas, K. Gubbins, W. Streett, and D. Tildesley, Mol. Phys. {\bf 37},
  1429  (1979).

\bibitem{Stell72}
G. Stell, J.~C. Rasaiah, and H. Narang, Mol. Phys. {\bf 23},  393  (1972).

\bibitem{Kusalik90}
P. Kusalik, J. Chem. Phys. {\bf 93},  3520  (1990).

\bibitem{Kusalik94}
P. Kusalik, Mol. Phys. {\bf 81},  199  (1994).

\bibitem{Deleeuw80}
S.~W. de~Leeuw, J.~W. Perram, and E.~R. Smith, Proc. R. Soc. Lond. A {\bf 373},
   27  (1980).

\bibitem{Allen87}
M. Allen and D. Tildesley, {\em Computer Simulation of Liquids} (Clarendon
  Press, Oxford, 1987).

\bibitem{Groh94}
B. Groh and S. Dietrich, Phys. Rev. Lett. {\bf 72},  2422  (1994);
Phys. Rev. {E50}, 3814 (1994).

\bibitem{Smit93}
B. Smit,  in {\em Computer Simulation in Chemical Physics} (Kluwer Academic
  Publishers, Dordrecht, 1993), pp.\ 173--209.

\bibitem{Panagiotopoulos92a}
A. Panagiotopoulos, Mol. Sim. {\bf 9},  1  (1992).

\bibitem{Powles82}
J. Powles, W. Evans, and N. Quirke, Mol. Phys. {\bf 46},  1347  (1982).

\bibitem{Smit90}
B. Smit and C. Williams, J. Phys. Condens. Matter {\bf 2},  4281  (1990).

\bibitem{Tsebers82}
A.~O. Tsebers, Magnetohydrodynamics {\bf 18},  137  (1982).

\bibitem{Zhang94}
H. Zhang and M. Widom, Phys. Rev. {\bf E49},  3591  (1994).

\bibitem{Sano83}
K. Sano and M. Doi, J. Phys. Soc. Jpn. {\bf 52},  2810  (1983).

\bibitem{Weis93b}
J. Weis and D. Levesque, Phys. Rev. {\bf E48},  3728  (1993).

\bibitem{Binder88}
K. Binder and D. Heermann, {\em Monte Carlo Simulation in Statistical Physics}
  (Springer Verlag, Berlin, 1988).

\bibitem{Grasselli94}
Y. Grasselli, G. Bossis, and E. Lemaire, J. Phys. II France {\bf 4},  253
  (1994).

\bibitem{Fermigier92}
M. Fermigier and A. Gast, J. Coll. Int. Sci. {\bf 154},  522  (1992).

\bibitem{Bacri82}
J.-C. Bacri, D. Salin, and R. Massart, J. Physique Lett. {\bf 43},  179
  (1982).

\bibitem{Liu93a}
J. Liu, T. Mou, and G.~A. Flores (unpublished).

\bibitem{Liu93b}
J. Liu, E. Lawrence, M. Ivey, and G.~A. Flores (unpublished).

\bibitem{Krueger80}
D. Krueger, IEEE Trans. Magn. {\bf 16},  251  (1980).

\end{thebibliography}
\end{document}